\documentclass[francais]{gretsi}

\usepackage[utf8]{inputenc}

\usepackage[T1]{fontenc}

\usepackage{amsmath, amssymb, amsfonts}

\usepackage{amsmath,amssymb,amsfonts}

\usepackage{graphicx, color, caption, subcaption}
\graphicspath{{fig/}}

\usepackage{mathtools}

\DeclarePairedDelimiterX{\infdivx}[2]{(}{)}{%
	#1\;\delimsize\|\;#2%
}

\usepackage{pgf}
\usepackage{bm}

\newcommand{\guillemets}[1]{``#1''}
\newcommand{\set}[1]{\left\{#1\right\}}

\newcommand{\myvector}[1]{\bm{#1}}

\newcommand{\mymatrix}[1]{\bm{#1}}
\newcommand{\entry}[1]{\left[{#1}\right]}
\newcommand{\hermitian}{H}

\newcommand{\R}{\mathbb{R}} 
\newcommand{\C}{\mathbb{C}} 
\newcommand{\E}{\mathrm{E}} 
\newcommand{\e}{\mathrm{e}} 

\newcommand{\eqdef}{\triangleq}

\newcommand{\rC}{\mathcal{C}}

\newcommand{\rN}{\mathcal{N}}

\newcommand{\intEnt}[2]{\set{#1, \dots, #2}}
\newcommand{\norm}[1]{\left\lVert#1\right\rVert}

\DeclareMathOperator{\diag}{\mathrm{diag}}
\DeclareMathOperator*{\vect}{\mathrm{vec}}

\DeclareUnicodeCharacter{2212}{-}
\begin{document}


\titre{Caractérisation d'une Source Diffuse à partir des Moments de sa Densité de Puissance en Tomographie SAR}

\auteurs{
  \auteur{Colin}{Cros}{colin.cros@isae-supaero.fr}{1}
  \auteur{Laurent}{Ferro-Famil}{laurent.ferro-famil@isae-supaero.fr}{1,2}
}

\affils{
  \affil{1}{ISAE-Supaero,
        10 avenue Marc Pélegrin, BP 54032,
        31055 Toulouse Cedex 4, France
  }
  \affil{2}{CESBIO,
        18 avenue Edouard Belin, BPI 2801,
        31401 Toulouse Cedex 9, France
  }
}


\resume{Ce papier présente une méthode pour l'estimation de caractéristiques d'une source diffuse à partir de mesures interférométriques dans le cadre de la tomographie SAR. La méthode proposée repose sur l'utilisation des moments centrés de la densité de réflectivité et n'utilise aucun modèle a priori. Les performances de la méthode sont discutées en fonction des paramètres du réseau d'antennes (résolution et ambigüité).}

\abstract{This paper presents a method for estimating the characteristics of a diffuse source from interferometric measurements in the context of SAR tomography. The proposed method is based on the use of central moments of the reflectivity density and does not use any a priori model. The method's performance is discussed as a function of antenna array parameters (resolution and ambiguity).}

\maketitle


\section{Introduction}

La tomographie par Radar à Synthèse d'Ouverture (SAR) permet de caractériser à grande échelle les environnements naturels.  Cette technique est particulièrement adaptée à l'étude et au suivi des forêts \cite{aghababaei2020forest}. Dans ce but, un nouveau satellite équipé d'un capteur radar en bande P ($\lambda \approx 70$ cm) sera lancé sous peu par la mission BIOMASS de l'ESA. La tomographie SAR utilise une série d'acquisitions radar 2D, prises selon des trajectoires légèrement décalées, afin de reconstruire la densité de réflectivité verticale du milieu \cite{gini2005multibaseline}. Cependant, le faible nombre de passages limite fortement la résolution des profils reconstruits et ne permet pas une reconstitution fidèle des milieux continus tels que les forêts. Des travaux récents \cite{bou2024tropical} ont montré que l'utilisation de modèles paramétriques, supposant une forme pour la densité de réflectivité, permet d'estimer la hauteur de la canopée des forêts tropicales à l'aide de mesures radar en bande P, mais ne peut discriminer la forme réelle de la densité. Pour pallier cette limite, nous proposons dans cet article une méthode non paramétrique se basant sur l'estimation des moments de la densité de réflectivité. La méthode repose sur les estimateurs COMET (\emph{COvariance Matching Estimation Techniques}) \cite{ottersten1998covariance} qui cherchent à reconstruire la matrice de covariance des mesures et offrent une alternative stable au maximum de vraisemblance. De plus, lorsque couplés au principe d'invariance étendu (EXIP), ils permettent d'estimer à moindre coût les caractéristiques d'une source diffuse \cite{besson2000decoupled, zoubir2006modified}, mais requièrent un a priori sur sa distribution. La tomographie SAR repose sur les techniques de traitement de réseaux d'antennes. Des méthodes utilisant les moments des distributions ont déjà été proposées pour ces réseaux afin d'estimer l'angle d'arrivée d'une source diffuse \cite{shahbazpanahi2004covariance} ou sa dispersion \cite{valaee1995parametric, shahbazpanahi2001distributed}.

\section{Modèle tomographique SAR}\label{sec: Modèle SAR}

\begin{figure}
	\centering
	\includegraphics{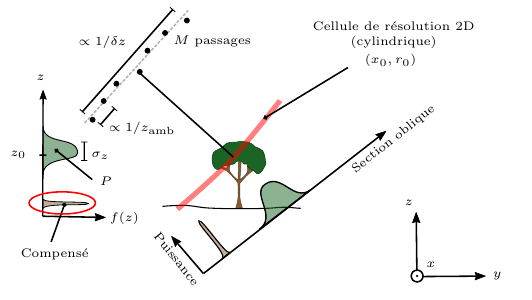}
	\caption{Principe de la tomographie SAR.}
	\label{fig: Illustration SAR tomography}
\end{figure}

Une mesure tomographique SAR est constituée de $M$ acquisitions SAR 2D prises selon des trajectoires légèrement décalées et idéalement parallèles, comme illustré sur la figure~\ref{fig: Illustration SAR tomography}. Sous l'hypothèse de Born, le signal réfléchi par la scène est la somme des contributions des différents réflecteurs. Le signal de la $m$-ième acquisition SAR 2D aux coordonnées (azimut, distance) $(x_0, r_0)$ se modélise comme :
\begin{equation}\label{eq: Modèle acquisition 2D}
	s_m(x_0, r_0) = \int h(x-x_0, r-r_0)a_c(\myvector{r})\e^{jk_cr}\, d\myvector{r} + \epsilon(x_0, r_0).
\end{equation}
Dans \eqref{eq: Modèle acquisition 2D}, $h(x,r)$ est la fonction d'ambigüité (2D), $k_c$ est le nombre d'onde bi-directionnel de la porteuse, $\myvector{r}$ désigne les coordonnées 3D du point, $r$ est sa distance à la trajectoire, $a_c(\myvector{r})$ désigne la densité de réflectivité cohérente de la scène, et $\epsilon(x_0,r_0)$ est le bruit d'acquisition supposé complexe circulaire gaussien. Pour les environnements naturels tels que les forêts, la densité de réflectivité est composée d'un grand nombre de contributions engendrées par des sources ponctuelles distribuées aléatoirement et suivant des distributions complexes circulaires gaussiennes. La densité vérifie alors : $\E[a_c(\myvector{r})] = 0$, $\E[a_c(\myvector{r}) a_c(\myvector{r} + \myvector{dr})] = 0$ et $\E[a_c(\myvector{r}) a_c(\myvector{r} + \myvector{dr})^*] = \sigma_a^2(\myvector{r}) \delta(\myvector{dr})$,
avec $\sigma_a^2(\myvector{r})$ désignant la densité de puissance de la réflectivité.

Les acquisitions SAR 2D présentent une ambigüité cylindrique autour de la trajectoire, puisque seuls la distance et l'azimut interviennent dans leur expression \eqref{eq: Modèle acquisition 2D}. Afin de lever cette ambiguïté et reconstruire la densité de réflectivité 3D, les $M$ acquisitions SAR 2D sont co-référencées sur un repère commun et démodulées en bande de base. L'image tomographique est alors un cube de données dans lequel chaque \guillemets{pixel}, $\myvector{y}(x_0, r_0)$, est un vecteur de $\C^M$. Ces vecteurs suivent une distribution complexe circulaire gaussienne, $\myvector{y} \sim \rC \rN(\myvector{0}, \mymatrix{R})$, dont la matrice de covariance $\mymatrix{R}$ est directement liée à la densité de réflectivité 3D de la scène. Pour les milieux homogènes sur le plan horizontal, la corrélation interférométrique entre deux acquisitions s'écrit :
\begin{equation}\label{eq: Corrélation interférométrique}
	\E[y_n y_m^*] = \int f(z) \e^{j (k_{z_{n}}-k_{z_{m}})z} \, dz + \sigma_\epsilon^2\delta(n-m).
\end{equation}
Dans \eqref{eq: Corrélation interférométrique}, $f(z)$ est la composante verticale de $\sigma_a^2(\myvector{r})$, $k_{z_{n}}$ désigne le nombre d'onde interférométrique de la $n$-ième acquisition, et les coordonnées de distance et d'azimut ont été omises pour plus de clarté.

La résolution de Fourier sur l'observation de $f(z)$ est $\delta z = 2\pi / (\max_n k_{z_n} - \min_m k_{z_m})$, et l'ambigüité s'approxime dans le cas d'un réseau quasi-uniforme par $z_\text{amb} \approx M \delta_z$. Ces valeurs sont proportionnelles à l'espacement entre les trajectoires. Pour l'observation d'une forêt, on dispose typiquement de $M = 7$ passages, leurs espacements sont choisis pour obtenir une ambigüité $z_\text{amb} = 100$~m, et donc une résolution $\delta z \approx 14$~m. La densité de réflectivité $f(z)$ est la somme des réponses de la forêt et du sol. Cependant, la réponse du sol peut être annulée à l'aide de prétraitements (\emph{notching}) \cite{mariotti2020notch}. On supposera donc que $f(z)$ ne contient que la réponse de la végétation.
On note $P$ la puissance réfléchie et $z_0$ le barycentre de sorte que $f(z) = P p(z-z_0)$, avec $\int p = 1$ et $\int zp(z)\, dz = 0$. La matrice de covariance $\mymatrix{R}$ s'écrit alors:
\begin{equation}
	\mymatrix{R} = \myvector{a}(z_0) \myvector{a}(z_0)^{\hermitian} \odot P \mymatrix{B} + \sigma_\epsilon^2 \mymatrix{I},
\end{equation}
où $\myvector{a}(z) = (\e^{jk_{z_1}z}, \dots, \e^{jk_{z_M}z})^{\intercal}$ est le vecteur directionnel et $\mymatrix{B}$ est une matrice de forme dont les coefficients valent $\entry{\mymatrix{B}}_{n,m} = \int p(z) \e^{j (k_{z_n}- k_{z_m})z} \, dz$.
 
\medbreak

L'objectif de l'étude est d'estimer trois caractéristiques de la densité de réflectivité : $(i)$ sa puissance $P$, $(ii)$ sa valeur moyenne $z_0$, et $(iii)$ sa largeur $\sigma_z^2 = \int z^2 p(z)\,dz$. Pour cela, on dispose d'une série de mesures $\set{\myvector{y}(t)}$ pour $t \in \intEnt{1}{N}$. La difficulté provient du fait que la forme de $f$ est inconnue. 

\section{Estimation à l'aide des moments}\label{sec: Méthode}

\subsection{Discussion sur l'observabilité}

\begin{figure*}[t]
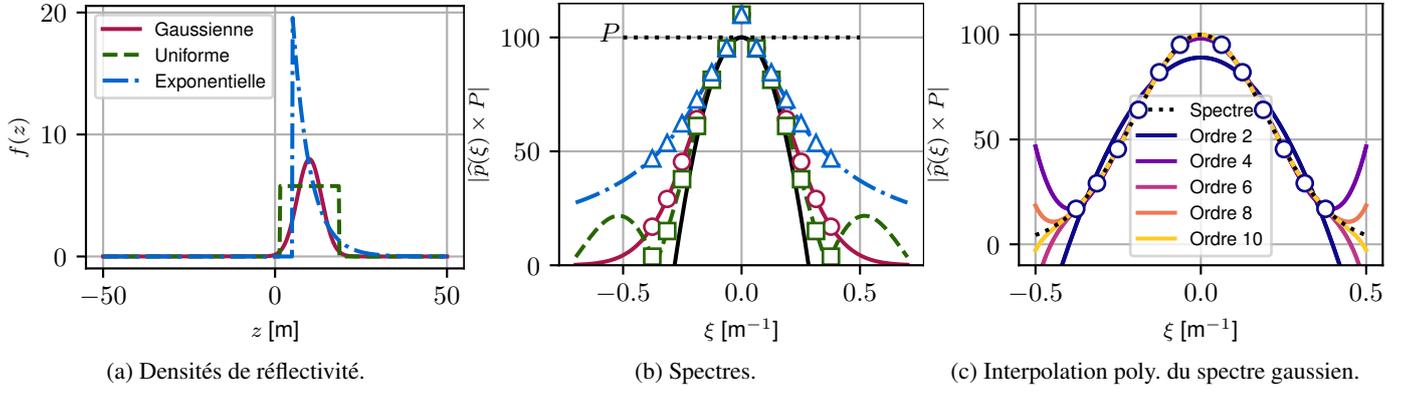

	\centering
	\null\hfill
	\subfloat[Densités de réflectivité.]{\input{fig/comparison_spectre_pdf.pgf}}
	\hfill
	\subfloat[Spectres.]{\input{fig/comparison_spectre_spectre.pgf}}
	\hfill
	\subfloat[Interpolation poly. du spectre gaussien.]{\input{fig/interpolation_spectre_s5.pgf}}
	\hfill\null
	\caption{Comparaison des spectres et des mesures disponibles pour différentes distributions ayant les mêmes statistiques. Les points de mesures sont représentés par les marqueurs sur les spectres. Sur la figure (b), la parabole $\xi \mapsto P - P\sigma_z^2 \xi^2 /2$ a également été tracée en noir. Les valeurs numériques utilisées sont $z_0 = 10$, $\sigma_z = 5$, $P = 100$, $\sigma_\epsilon^2 = 10$, $M = 7$ et $z_\text{amb} = 100$.} 
	\label{fig: Comparaison spectres}
\end{figure*}

Si la distribution $p$ était connue, on pourrait construire les estimateurs du maximum de vraisemblance pour les trois caractéristiques. Malheureusement, ce n'est pas le cas et la densité de réflectivité ne suit aucune loi particulière. De plus, supposer une loi particulière incorrecte pour $f$, aboutit généralement à des estimateurs biaisés, l'apparition de ces biais est illustrée dans la section~\ref{sec: Simulations}. Nous recherchons par conséquent un modèle non paramétrique.

Les mesures permettent de construire une estimation de la matrice de covariance $\mymatrix{R}$ comme :
\begin{equation}
	\mymatrix{\bar{R}} = \frac{1}{N} \sum_t \myvector{y}(t)\myvector{y}(t)^{\hermitian}.
\end{equation}
En introduisant la distribution unitaire centrée des réflecteurs $p(z) = f(z-z_0)/P$, et sa transformée de Fourier $\widehat p(\xi)$, les coefficients de la matrice de covariance sont :
\begin{equation}
	\entry{\mymatrix{R}}_{n,m} = \left\{\begin{array}{cl}
		P + \sigma_\epsilon^2 & \text{si $n = m$}, \\
		P \widehat{p}(k_{z_m} - k_{z_n}) \e^{j(k_{z_m}-k_{z_n})z_0} & \text{sinon}.
	\end{array}\right.
\end{equation}
Les mesures réalisées permettent d'observer le spectre de la distribution de réflecteurs. À partir des coefficients de $\mymatrix{R}$, on ne peut reconstruire au mieux qu'une version passe-bas de la densité de réflectivité. De plus, les coefficients de $\mymatrix{R}$ correspondent à l'évaluation de la fonction $\xi \mapsto P \widehat p(\xi) \e^{j\xi z_0}$, sauf pour les coefficients diagonaux qui sont corrompus par le bruit de mesure. La puissance $P$ correspond à la valeur (manquante) en $0$ et la covariance spatiale $\sigma_z^2$ à la courbure de $\widehat p$ en $0$. Par conséquent, ces deux caractéristiques peuvent être estimées en réalisant une interpolation polynomiale des mesures. La figure~\ref{fig: Comparaison spectres} présente différentes distributions potentielles ainsi que les observations de leurs spectres. On constate qu'un développement limité au deuxième ordre fournit déjà une excellente approximation.

\subsection{Estimateur COMET}

L'estimation des caractéristiques de la forêt est réalisée à l'aide d'un estimateur COMET. Ces estimateurs cherchent les paramètres reconstituant au mieux, au sens des moindres carrés, la matrice de covariance empirique. De par leur stabilité et leurs performances asymptotiques, ils constituent une alternative au maximum de vraisemblance. Si la matrice de covariance est paramétrée par un vecteur $\myvector{\theta}$ comme $\mymatrix{\hat R}(\myvector{\theta})$, alors l'estimateur COMET s'obtient en minimisant la fonction de coût suivante :
\begin{equation}\label{eq: Fonction de coût}
	L(\myvector{\theta}) = \norm{\mymatrix{W}^{\hermitian/2}(\mymatrix{\bar {R}}- \mymatrix{\hat R}(\myvector{\theta})) \mymatrix{W}^{1/2}}^2,
\end{equation}
où $\mymatrix{W}$ est une matrice de pondération. Deux choix sont couramment pris, $\mymatrix{W} = \mymatrix{I}$ qui correspond à une méthode des moindres carrés non pondérés, et $\mymatrix{W} = \mymatrix{\bar{R}}^{-1}$ qui engendre un estimateur efficient si le modèle $\mymatrix{\hat R}(\myvector{\theta})$ est bien spécifié \cite{ottersten1998covariance}.

Compte tenu de la méconnaissance de la forme de la distribution, nous choisissons de la modéliser à partir de ses moments centrés : $\mu_d = \int z^d p(z) \, dz$. Ce choix est motivé par le fait que si l'étendue du volume est petite, alors la transformée de Fourier $\widehat p$ sera très plate et pourra être approximée par les premiers termes de son développement limité, qui correspondent aux moments de $p$. On notera que par définition de $p$, $\mu_0 = 1$ et $\mu_1 = 0$. Pour un vecteur $\myvector{\mu} = \begin{pmatrix} \mu_2, \dots, \mu_D \end{pmatrix} \in \R^{D-1}$, on approxime alors la fonction $\widehat p$ par :
\begin{equation}
	\widehat p(\xi, \mu) = 1 + \sum_{d=2}^D \frac{j^d}{d!}\mu_d \xi^d.
\end{equation}
En introduisant le vecteur de paramètre $\myvector{\theta} = (z_0, P, \sigma_\epsilon^2, \myvector{\mu})$, la matrice de forme $\mymatrix{B}$ et la matrice de covariance s'approximent alors par :
\begin{subequations}
	\begin{align}
		\mymatrix{\hat{B}}(\mu) &= \myvector{1}\myvector{1}^{\intercal} + \sum_{d=2}^D \frac{j^d}{d!}\mu_d \mymatrix{U}^{(d)}, \\
		\mymatrix{\hat R}(\myvector{\theta}) &= \myvector{a}(z_0) \myvector{a}(z_0)^{\hermitian} \odot P \mymatrix{\hat{B}}(\myvector{\mu}) + \sigma_\epsilon^2 \mymatrix{I},
	\end{align}
\end{subequations}
où les matrices $\mymatrix{U}^{(d)}$ ne dépendent que de la géométrie du système, leurs coefficients valent : $\entry{\mymatrix{U}^{(d)}}_{n,m} = (k_{z_n} - k_{z_m})^d$.

Avec la modélisation proposée, les moments entrent linéairement dans l'expression des coefficients de la matrice $\mymatrix{\hat{B}}(\myvector{\theta})$. De même, avec le changement de variable $\myvector{\nu} = P \myvector{\mu}$ et la nouvelle paramétrisation $\myvector{\theta} = (z_0, P, \sigma_\epsilon^2, \myvector{\nu})$, tous les paramètres sauf $z_0$ entrent également linéairement dans l'expression des coefficients. Notons $\myvector{\alpha} = (P, \sigma_\epsilon^2, \myvector{\nu}) \in \R^{D+1}$ l'ensemble de ces paramètres linéaires. La fonction de coût \eqref{eq: Fonction de coût}, se ré-écrit :
\begin{equation}
	L(z_0, \myvector{\alpha}) = (\myvector{\bar{r}} - \mymatrix{\Phi}(z_0)\mymatrix{J} \myvector{\alpha})^{\hermitian}
	(\mymatrix{W}^{\intercal}\otimes\mymatrix{W})
	(\myvector{\bar{r}} - \mymatrix{\Phi}(z_0)\mymatrix{J} \myvector{\alpha}),
\end{equation}
où $\myvector{\bar{r}} = \vect{\mymatrix{\bar{R}}}$ est le vecteur obtenu en concaténant les colonnes de la matrice, $\mymatrix{J}$ est la matrice telle que $\mymatrix{J}\myvector{\alpha} = \vect\left\{P\mymatrix{B}(\myvector{\mu})\right\} + \sigma_\epsilon^2 \mymatrix{I}$ et :
\begin{align*}
	\mymatrix{\Phi}(z_0) &= \diag \myvector{a}(z_0), & \mymatrix{\Psi}(z_0) &= \mymatrix{\Phi}(z_0)^{\hermitian} \otimes \mymatrix{\Phi}(z_0).
\end{align*}

Comme la fonction de coût est quadratique, pour un $z_0$ fixé, le vecteur $\myvector{\alpha}$ minimisant le coût $L(z_0, \bullet)$ se calcule de manière explicite :
\begin{multline}\label{eq: Solution alpha}
	\myvector{\hat \alpha}(z_0) = \left[(\mymatrix{\Psi}(z_0) \mymatrix{J})^{\hermitian} (\mymatrix{W}^{\intercal}\otimes\mymatrix{W})(\mymatrix{\Psi}(z_0)\mymatrix{J})\right]^{-1}\\
	\times \mymatrix{\Psi}(z_0) \mymatrix{J})^{\hermitian} (\mymatrix{W}^{\intercal}\otimes\mymatrix{W})\myvector{\bar r}.
\end{multline}

L'optimisation de $L$ ne requiert alors qu'une recherche sur $z_0$, paramètre contraint sur un segment compte tenu de l'ambigüité du système. On a alors :
\begin{equation}\label{eq: Solution omega}
	\hat z_0 = \arg \max \myvector{y}(z)^{\hermitian} \mymatrix{Y}(z)^{-1} \myvector{y}(z),
\end{equation}
avec :
\begin{align*}
	\myvector{y}(z) &\eqdef (\mymatrix{\Psi}(z) \mymatrix{J})^{\hermitian} (\mymatrix{W}^{\intercal}\otimes\mymatrix{W})\myvector{\bar r}, \\
	\mymatrix{Y}(z) &\eqdef (\mymatrix{\Psi}(z) \mymatrix{J})^{\hermitian} (\mymatrix{W}^{\intercal}\otimes\mymatrix{W})\mymatrix{\Psi}(z)\mymatrix{J}.
\end{align*}
Le paramètre $\myvector{\hat \alpha}$ optimal s'obtient ensuite en réinjectant la solution de \eqref{eq: Solution omega} dans \eqref{eq: Solution alpha}.

\section{Simulations et discussion}\label{sec: Simulations}

\begin{figure*}[t]
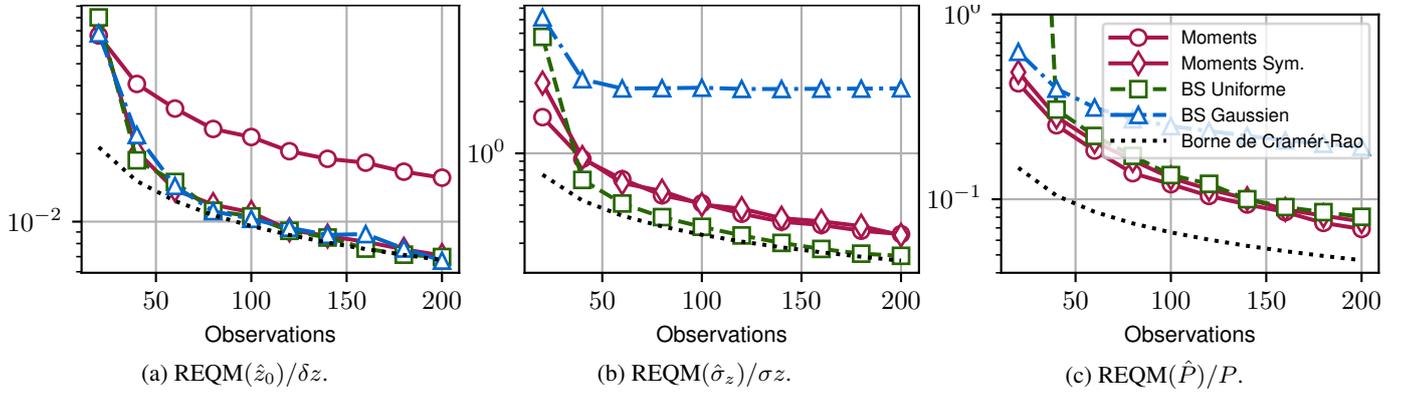

	\centering
	\null\hfill
	\subfloat[$\text{REQM}(\hat z_0)/\delta z$.]{\input{fig/comparison_MSE_z0.pgf}}
	\hfill
	\subfloat[$\text{REQM}(\hat \sigma_z)/\sigma z$.]{\input{fig/comparison_MSE_sigma_z.pgf}}
	\hfill
	\subfloat[$\text{REQM}(\hat P)/P$.]{\input{fig/comparison_MSE_P.pgf}}
	\hfill\null
	\caption{Comparaison des erreurs d'estimation en fonction du nombre d'observations $N$ pour une distribution uniforme. Les REQM ont été calculées à partir de $5\,000$ réalisations. Les mêmes valeurs numériques que pour la fig.~\ref{fig: Illustration SAR tomography} ont été utilisées.}
	\label{fig: Comparaison erreurs}
\end{figure*}

\begin{figure*}[t]
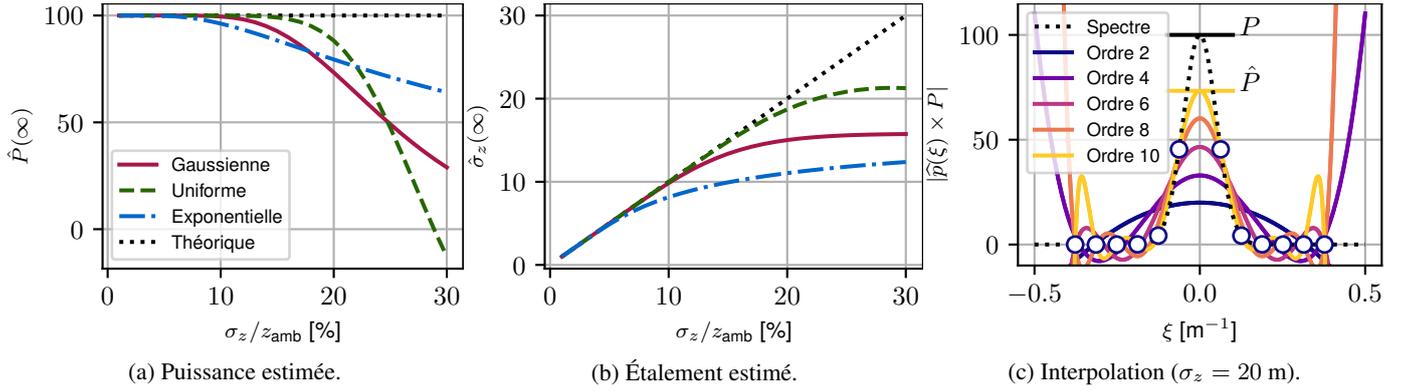

	\centering
	\null\hfill
	\subfloat[Puissance estimée.]{\input{fig/comparison_biais_P.pgf}}
	\hfill
	\subfloat[Étalement estimé.]{\input{fig/comparison_biais_sigma_z.pgf}}
	\hfill
	\subfloat[Interpolation ($\sigma_z = 20$ m).]{\input{fig/interpolation_spectre_s20.pgf}}
	\hfill\null
	\caption{Comparaison des estimations asymptotiques ($N \to \infty$) en fonction de l'étalement $\sigma_z$. Les mêmes valeurs numériques que pour la fig.~\ref{fig: Illustration SAR tomography} ont été utilisées.}
	\label{fig: Biais asymptotiques}
\end{figure*}

La méthode de moments est comparée à la méthode paramétrique proposée par Besson et Stoica  \cite{besson2000decoupled} reposant également sur un estimateur COMET et utilisant un modèle pour la forme de la distribution. La figure~\ref{fig: Comparaison erreurs} présente l'évolution des racines de l'erreur quadratique moyenne (REQM) pour les deux méthodes. La méthode des moments a été appliquée telle que décrite dans la section~\ref{sec: Méthode} et adaptée afin d'imposer la symétrie de la distribution (en n'utilisant que les moments pairs). La méthode paramétrique de Besson et Stoica (BS) a été appliquée en supposant la bonne distribution (uniforme) et une distribution erronée (gaussienne). La borne de Cramér-Rao a également été tracée. On constate que la méthode paramétrique avec le bon modèle fournit les meilleurs résultats comme attendu, en revanche avec un modèle incorrect, il y a un biais significatif sur les estimations de l'étalement $\hat \sigma_z$ et de la puissance $\hat P$. La méthode des moments fournit quant à elle de bonnes estimations, on note qu'imposer la symétrie permet d'améliorer significativement la précision sur $\hat z_0$ (car la distribution réelle est symétrique).

La méthode des moments repose sur une interpolation polynomiale du spectre de la distribution. Si la distribution est trop étalée, son spectre est étroit et les points de mesures ne permettent plus de retrouver les caractéristiques de la sources. Dans ce cas, des biais apparaissent sur les estimations de $\hat \sigma$ et $\hat P$. La figure~\ref{fig: Biais asymptotiques} présente l'évolution de ces biais en fonction de l'étalement de la source ainsi que l'illustration des interpolations polynomiales obtenues dans le cas d'une source étendue ($\sigma_z / z_{\text{amb}} = 20$~\%). On observe l'apparition de biais dès que le rapport $\sigma_z / z_{\text{amb}}$ dépasse les $10$~\%. Pour les valeurs usuelles en tomographie SAR $\sigma_z \approx 5$~m et $z_{\text{amb}} \approx 100$~m, les biais sont négligeables. Ce phénomène illustre un compromis à trouver sur le choix de l'ambigüité $z_{\text{amb}}$ lors de la conception du système. Une forte ambigüité permet d'identifier avec une meilleure précision la puissance $P$ et l'étalement $\sigma_z$, en revanche cela se fait au détriment de la résolution, $\delta z \approx z_{\text{amb}} / M$, et donc de la précision de l'estimation de la hauteur $z_0$.

\section{Conclusion}\label{sec: Conclusion}

Ce papier a proposé une méthode basée sur les moments pour estimer les caractéristiques d'une forêt à l'aide de mesures interférométriques SAR. L'intérêt principal est de ne supposer aucun modèle a priori pour la densité de réflectivité observée qui ne suit en pratique aucune distribution simple. La méthode repose sur une interpolation polynomiale et ne fonctionne que si la résolution fréquentielle est suffisamment bonne, c'est-à-dire si l'ambigüité est grande devant le phénomène observé.

Les prochains travaux s'intéresseront aux calculs précis des bornes mal-spécifiées sur les variances des erreurs d'estimation.


\bibliography{bibliography}

\begin{thebibliography}{10}
\providecommand{\url}[1]{#1}
\csname url@samestyle\endcsname
\providecommand{\newblock}{\relax}
\providecommand{\bibinfo}[2]{#2}
\providecommand{\BIBentrySTDinterwordspacing}{\spaceskip=0pt\relax}
\providecommand{\BIBentryALTinterwordstretchfactor}{4}
\providecommand{\BIBentryALTinterwordspacing}{\spaceskip=\fontdimen2\font plus
\BIBentryALTinterwordstretchfactor\fontdimen3\font minus
  \fontdimen4\font\relax}
\providecommand{\BIBforeignlanguage}[2]{{%
\expandafter\ifx\csname l@#1\endcsname\relax
\typeout{** WARNING: IEEEtran.bst: No hyphenation pattern has been}%
\typeout{** loaded for the language `#1'. Using the pattern for}%
\typeout{** the default language instead.}%
\else
\language=\csname l@#1\endcsname
\fi
#2}}
\providecommand{\BIBdecl}{\relax}
\BIBdecl

\bibitem{aghababaei2020forest}
H.~Aghababaei, G.~Ferraioli, L.~Ferro-Famil, Y.~Huang,
  M.~Mariotti~D'Alessandro, V.~Pascazio, G.~Schirinzi, and S.~Tebaldini,
  ``Forest {SAR} tomography: Principles and applications,'' \emph{IEEE
  Geoscience and Remote Sensing Magazine}, vol.~8, no.~2, pp. 30--45, 2020.

\bibitem{gini2005multibaseline}
F.~Gini and F.~Lombardini, ``Multibaseline cross-track {SAR} interferometry: a
  signal processing perspective,'' \emph{IEEE Aerospace and Electronic Systems
  Magazine}, vol.~20, no.~8, pp. 71--93, 2005.

\bibitem{bou2024tropical}
P.-A. Bou, L.~Ferro-Famil, F.~Brigui, and Y.~Huang, ``Tropical forest
  characterisation using parametric {SAR} tomography at {P} band and
  low-dimensional models,'' \emph{IEEE Geoscience and Remote Sensing Letters},
  pp. 1--1, 2024.

\bibitem{ottersten1998covariance}
O.~Ottersten, P.~Stoica, and R.~Roy, ``Covariance matching estimation
  techniques for array signal processing applications,'' \emph{Digital Signal
  Processing}, vol.~8, no.~3, pp. 185--210, 1998.

\bibitem{besson2000decoupled}
O.~Besson and P.~Stoica, ``Decoupled estimation of {DOA} and angular spread for
  a spatially distributed source,'' \emph{IEEE Transactions on Signal
  Processing}, vol.~48, no.~7, pp. 1872--1882, 2000.

\bibitem{zoubir2006modified}
A.~Zoubir, Y.~Wang, and P.~Chargé, ``A modified {COMET-EXIP} method for
  estimating a scattered source,'' \emph{Signal Processing}, vol.~86, no.~4,
  pp. 733--743, 2006.

\bibitem{shahbazpanahi2004covariance}
S.~Shahbazpanahi, S.~Valaee, and A.~Gershman, ``A covariance fitting approach
  to parametric localization of multiple incoherently distributed sources,''
  \emph{IEEE Transactions on Signal Processing}, vol.~52, no.~3, pp. 592--600,
  2004.

\bibitem{valaee1995parametric}
S.~Valaee, B.~Champagne, and P.~Kabal, ``Parametric localization of distributed
  sources,'' \emph{IEEE Transactions on Signal Processing}, vol.~43, no.~9, pp.
  2144--2153, 1995.

\bibitem{shahbazpanahi2001distributed}
S.~Shahbazpanahi, S.~Valaee, and M.~Bastani, ``Distributed source localization
  using {ESPRIT} algorithm,'' \emph{IEEE Transactions on Signal Processing},
  vol.~49, no.~10, pp. 2169--2178, 2001.

\bibitem{mariotti2020notch}
M.~Mariotti~d{’}Alessandro, S.~Tebaldini, S.~Quegan, M.~J. Soja, L.~M.~H.
  Ulander, and K.~Scipal, ``Interferometric ground cancellation for above
  ground biomass estimation,'' \emph{IEEE Transactions on Geoscience and Remote
  Sensing}, vol.~58, no.~9, pp. 6410--6419, 2020.

\end{thebibliography}


\end{document}